\documentclass[12pt]{article}
\usepackage{graphicx}
 \usepackage{amsmath}
\usepackage{amssymb} 
\usepackage{epsfig}
\usepackage{color}
\usepackage{latexsym}
\usepackage{amsmath}
\usepackage{amssymb}
\usepackage{dsfont}
\usepackage{verbatim}  

\newcommand\pubnumber{JLAB-THY-15-2126}
\newcommand\pubdate{\today}

\def\wayne{Department of Physics, University of Ljubljana, Jadranska 19, Ljubljana, Slovenia\\
Jozef Stefan Institute, Jamova 39, 1000 Ljubljana, Slovenia\\
Theory Center, Jefferson Lab, 12000 Jefferson Avenue, Newport News, VA 23606, USA }
\def\support{\footnote{ sasa.prelovsek@ijs.si}}

\def\Title#1{\begin{center} {\Large #1 } \end{center}}
\def\Author#1{\begin{center}{ \sc #1} \end{center}}
\def\Address#1{\begin{center}{ \it #1} \end{center}}

\newcommand\pubblock{\rightline{\begin{tabular}{l} \pubnumber\\
         \pubdate  \end{tabular}}}
\newenvironment{Abstract}{\begin{quotation}  }{\end{quotation}}
\newenvironment{Presented}{\begin{quotation} \begin{center} 
             PRESENTED AT\end{center}\bigskip 
      \begin{center}\begin{large}}{\end{large}\end{center} \end{quotation}}
\def\Acknowledgements{\bigskip  \bigskip \begin{center} \begin{large}
             \bf ACKNOWLEDGEMENTS \end{large}\end{center}}

\def\beq{\begin{equation}}
\def\eeq#1{\label{#1}\end{equation}}
\def\eeqn{\end{equation}}

\def\beqa{\begin{eqnarray}}
\def\eeqa#1{\label{#1}\end{eqnarray}}
\def\eeqan{\end{eqnarray}}

\let\bar=\overbar

\def\Dslash{\not{\hbox{\kern-4pt $D$}}}
\def\dslash{\not{\hbox{\kern-2pt $\del$}}}

\def\msb{{\bar{\ssstyle M \kern -1pt S}}}

\begin{document}
\begin{titlepage}
\pubblock

\vfill
\Title{Lattice studies of charmonia and exotics }
\vfill
\Author{Sasa Prelovsek\support}
\Address{\wayne}
\vfill
\begin{Abstract}
 The lattice QCD simulations of charmonia and exotic charmonium-like states are reviewed. 
 I report on the first exploratory simulation of  charmonium resonances above open charm threshold  which takes into account the strong decay.  The puzzles related to the first-excited scalar charmonia are discussed. Evidence for $X(3872)$ is  presented along with investigation of its Fock components. The $Z_c^+(3900)$ seems to emerge from the HALQCD approach as a result of the  coupled channel effect $J/\psi \pi -D\bar D^*$.   The indication for the pentaquark bound state $\eta_c N$ is presented. 
 \end{Abstract}
\vfill
\begin{Presented}
The 7th International Workshop on Charm Physics (CHARM 2015)\\
Detroit, MI, 18-22 May, 2015
\end{Presented}
\vfill
\end{titlepage}
\def\thefootnote{\fnsymbol{footnote}}
\setcounter{footnote}{0}
%

\section{Introduction} 

Experiments discovered   a number of interesting charmonium-like states which can not be reconciled with conventional $\bar  cc$. The prominent examples are tetraquarks $Z_c^+$ with flavour content $\bar cc\bar d u$ and pentquarks $P_c$ with  $\bar cc uud$ \cite{Aaij:2015tga}. There is no doubt that establishing  the conventional and the exotic charmonia from ab-initio lattice QCD simulations is desired.  I review  state of art calculations  along these lines. 

\section{Lattice methodology}

The physics information on a hadron (below, near or above threshold) is commonly extracted from the discrete energy spectrum in lattice QCD. 
The physical system for given quantum numbers is created from the vacuum $|\Omega\rangle$ using interpolator ${\cal O}_j^\dagger$  at time $t\!=\!0$ and the system propagates for time $t$ before being annihilated by ${\cal O}_i$.    To study a charmonium or a charmonium-like state with given $J^P$ one  can, for example,  use ${\cal O}\simeq \bar c	\Gamma c,~$ two-meson interpolators ${\cal O}=(\bar c \Gamma_1  q)(\bar q \Gamma_2  c),~$$(\bar c \Gamma_1  c)(\bar q \Gamma_2  q)~$ and  ${\cal O}=[\bar c \Gamma_1 \bar q][c\Gamma_2 q]$ with desired quantum numbers.  After the spectral decomposition the correlators are expressed in terms of the energies  $E_n$ of eigenstates $|n\rangle$ and their overlaps $Z_j^n$
\begin{equation}
\label{C}
C_{ij}(t)= \langle \Omega|{\cal O}_i (t) {\cal O}_j^\dagger (0)|\Omega \rangle=\sum_{n}Z_i^nZ_j^{n*}~e^{-E_n t}~,\qquad Z_i^n\equiv \langle \Omega|{\cal O}_i|n\rangle~.
\end{equation}
The correlators are evaluated on the lattice and their time-dependence allows to extract $E_n$ and $Z_n^i$  \cite{Michael:1985ne,Blossier:2009kd}. 

 The energy eigenstates $|n\rangle$ are   predominantly "one-meson" states (e.g. $\chi_{c1}(1P)$) or predominantly "two-meson" states (e.g. $D\bar D^*$) - in interacting theory they are mixtures of those.    Two-meson states  have a discrete spectrum due to the periodic boundary condition on the finite lattices.  If they do not  interact, then the momenta of each meson is $\vec{p}= \!\tfrac{2\pi}{L}\vec{N}$ with $\vec{N}\in {N}^3$, and the non-interacting energies of $M_1(\vec p)M_2(-\vec p)$    are $E^{n.i.}=E_1(p)+E_2(p)$ with $E_{1,2}(p)=(m_{1,2}^2+p^2)^{1/2}$.  The energies $E_n$ extracted from the lattice  are slightly shifted in presence of the interaction and the shift  provides rigorous information on the scattering matrix, as discussed in  Section \ref{sec:scat}. 
In experiment, two-meson  states correspond to the two-meson decay products with a continuous energy spectrum. 

The Wick contractions for charmonium(like) states contain  terms where charm quark propagates from source to sink, and terms where charm quark annihilates. The charm quark annihilation is omitted in all lattice simulations reported in this talk. One expects very small influence from charm annihilation on the energies of eigenstates of interest, but this needs to be verified in the future. It will be very challenging to go beyond this approximation 
     due to a large number of light single and multi-hadron states with the same quantum numbers.
For some channels (for example $D\bar D^*$ scattering with $I=0$), there are also Wick contractions where $u,d,s$ annihilate and these Wick contractions have been consistently taken into account in all the described simulations.

  \begin{figure}[htb]
\centering 
\includegraphics[width=0.45\textwidth,clip]{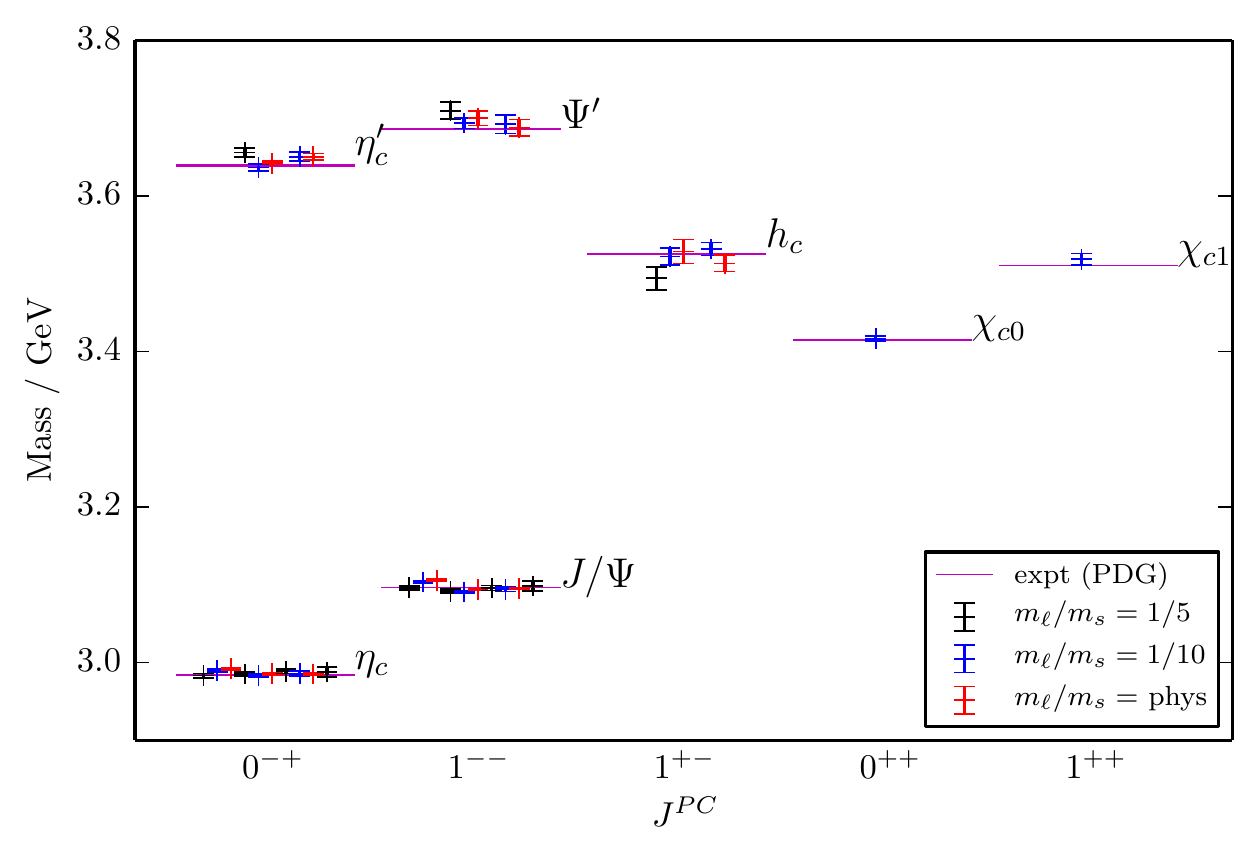} 
\includegraphics[width=0.5\textwidth,clip]{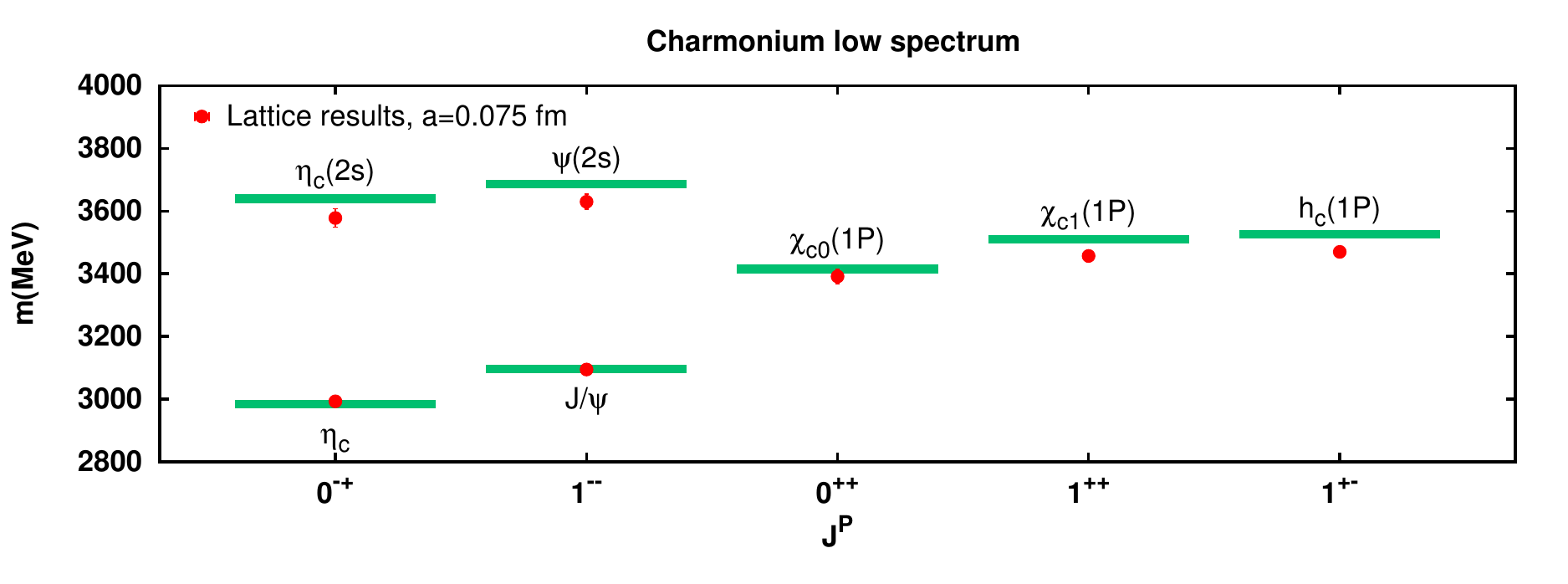} 
\caption{ Spectra of charmonia below open charm threshold \cite{Galloway:2014tta} (left) and  \cite{Bali:2015lka} (right).}
\label{fig:below_th}
\end{figure}

 \section{Charmonia well below open charm threshold}
 
  The masses $m_i=E_i(P=0)$ are extracted  from the energies obtained with $\bar cc$ interpolating fields,  which are extrapolated $a\to 0$, $V\to \infty$ and $m_q\to m_q^{phys}$. Since charm annihilation is omitted, there is no complication arising in relation to the multi-hadron states. 
 Some of the recent spectra \cite{Yang:2014sea,Galloway:2014tta,Bali:2015lka,Mohler:2014ksa} are shown in Fig. \ref{fig:below_th}.   The results are in impressive agreement with experimental masses and the main remaining uncertainty is due to the omission of charm annihilation. 
 
  \section{Excited charmonia within single-meson approach}
  
  The most extensive spectra of the excited charmonia have been calculated within the so-called single-meson approach  Hadron Spectrum Collaboration in 2012 \cite{Liu:2012ze}.  The continuum $J^{PC}$ was reliably identified using the advanced spin-identification method.  Multiplets of hybrid states were also found and some of them carry exotic $J^{PC}$.  
  The single-meson treatment   ignores strong decays of resonances and threshold effects.  It gives valuable reference spectra, but can not give reliable conclusions on  the near-threshold exotic states, for example. 
  
    \begin{figure}[htb] 
\begin{center}
\includegraphics[width=0.4\textwidth,clip]{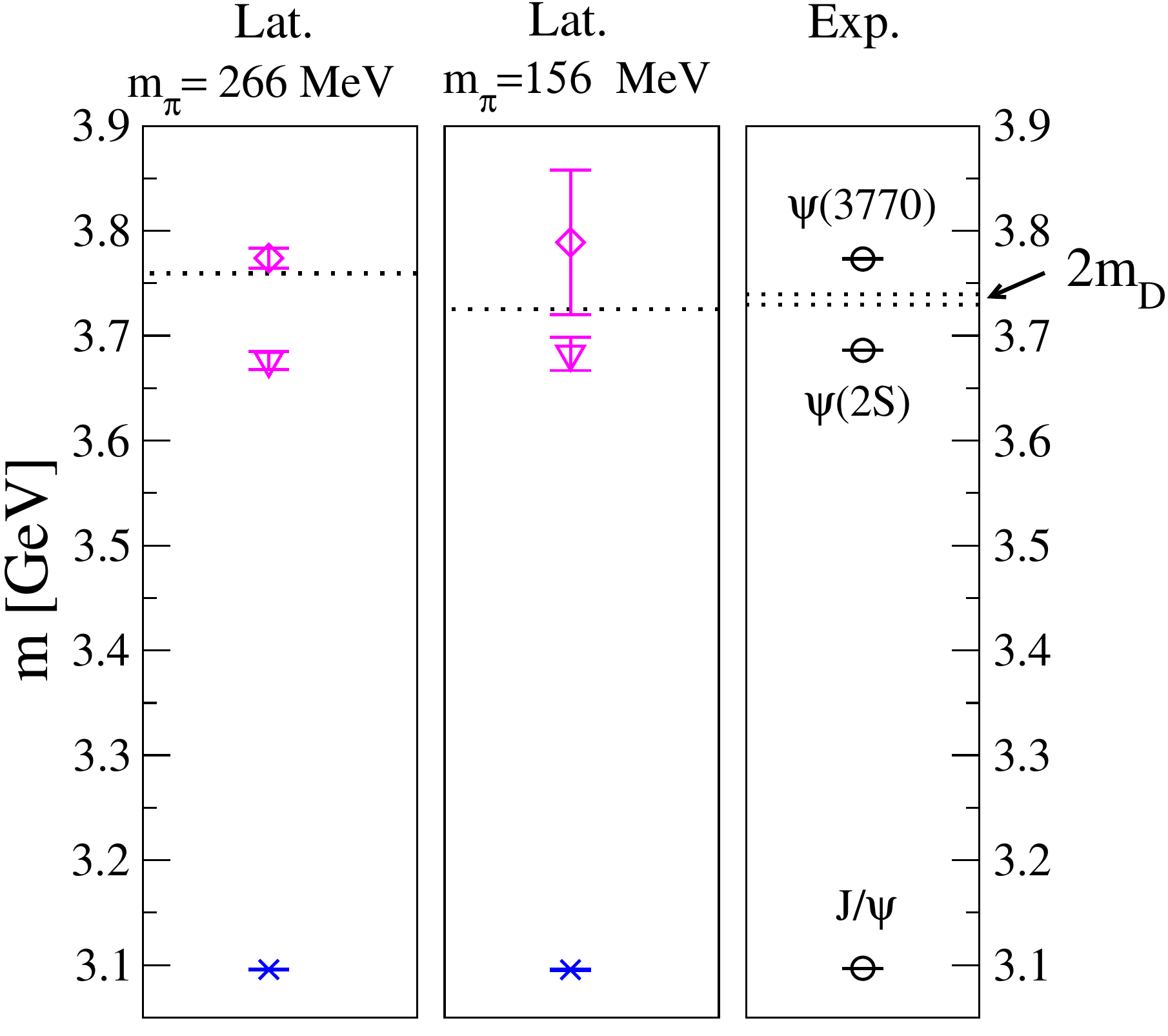} $\quad$ 
\includegraphics[width=0.49\textwidth,clip]{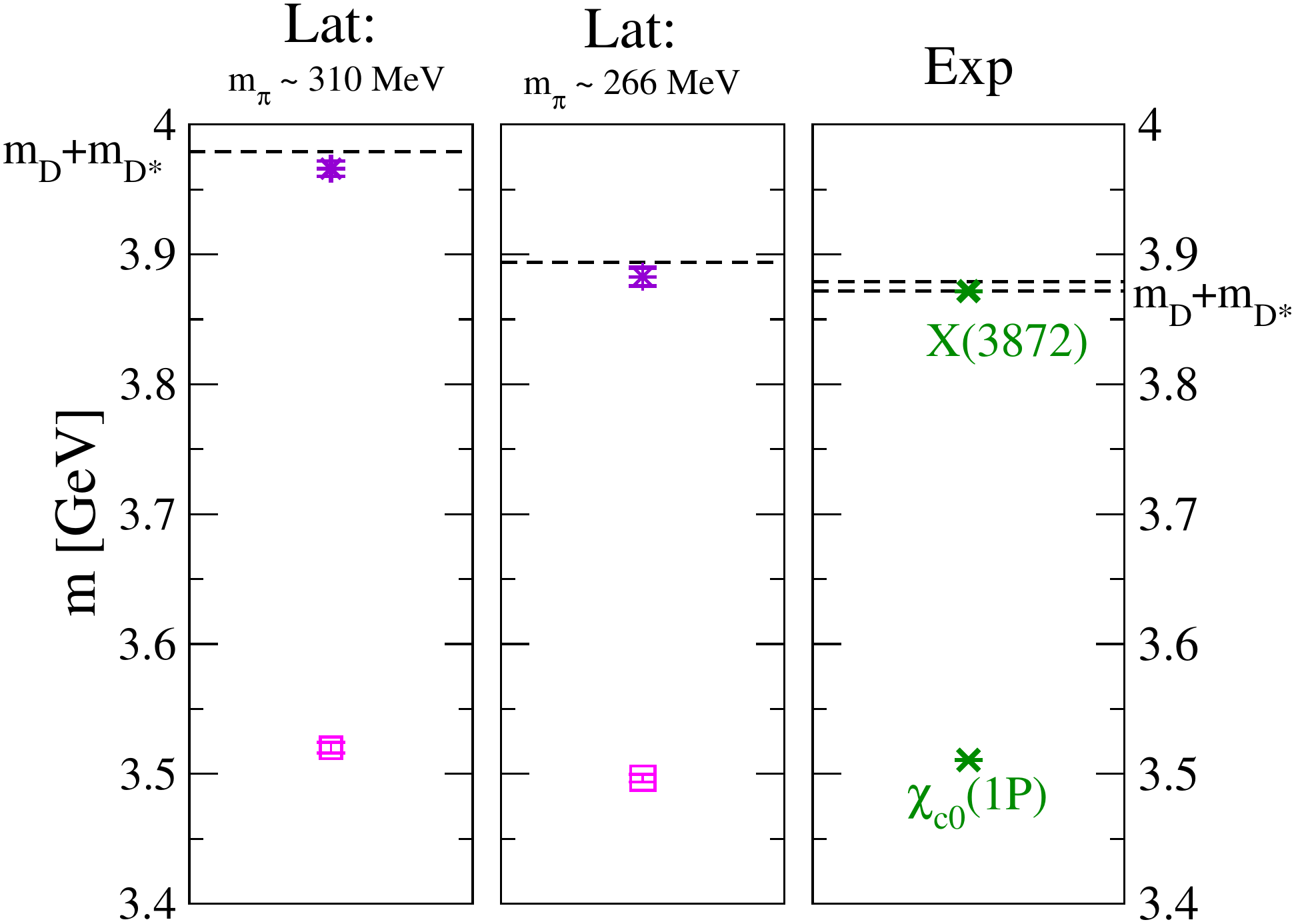}  
\caption{\label{fig:vector_charmonia}   Left: The  spectrum of the vector charmonia from \cite{Lang:2015sba}: the  diamond  denotes the resonance mass of  $\psi(3770)$, while the triangle denotes the pole mass of the bound state $\psi(2S)$; both are obtained from $D\bar D$ scattering matrix.   Right: The location of $X(3872)$ with $I=0$ which emerges as shallow bound state in  $D\bar D^*$ scattering at $m_\pi\simeq 266~$MeV  \cite{Prelovsek:2013cra} and  $m_\pi\simeq 310~$MeV \cite{Lee:2014uta_private} (update of  \cite{Lee:2014uta}).  }
\end{center}
\end{figure}

  \section{Scattering approach to resonances and bound states }\label{sec:scat}

   In the energy region near or above threshold, the masses of bound-states and resonances have to be inferred from the infinite-volume scattering matrix of the one-channel (elastic) or multiple-channel (inelastic) scattering.  The simplest example is a one-channel elastic scattering  in partial wave $l$, where   the scattering matrix  is parametrized in terms of the phase shift $\delta_l(p)$ and satisfies unitarity $SS^\dagger=1$
\begin{equation}
\label{T}
S(E)=e^{2i\delta_l(E)}\ , \quad S(E)=1+2iT(E)\ , \quad T(E)=\frac{1}{\cot(\delta_l(E))-i}~.
\end{equation}  
  L\"uscher has shown that the energy  $E$ of two-meson eigenstate in finite volume $L$ gives the elastic phase shift $\delta(E)$ at that energy in infinite volume  \cite{Luscher:1991cf}.  This relation and 
 its generalizations are at the core of extracting rigorous  information about the scattering from the recent lattice simulation.  It leads  $\delta(E)$ only for specific values of $E$ since  
  since spectrum  of two-meson eigenstates is discrete.  The $\delta(E)$ or $T(E)$ (\ref{T})  provide the masses of resonances and bound states:  
  \begin{itemize}
\item  In the vicinity of a {\it hadronic resonance} with mass $m_R$ and width $\Gamma$, the  cross section $\sigma \propto |T(p)|^2$ has a Breit-Wigner-type shape with  $\delta(s=m_R^2)=\tfrac{\pi}{2}$
\begin{equation}
\label{R}
T(p)=\frac{-\sqrt{s}~ \Gamma(p)}{s-m_R^2+i \sqrt{s}\, \Gamma(p)}=\frac{1}{\cot\delta(p)-i}\ , \quad
\Gamma(p)=g^2\,\frac{p^{2l+1}}{s} . 
\end{equation}  
The fit of $\delta_l(p)$  renders $m_R$ and $g$ or $\Gamma$. It is customary to compare $g$ rather than $\Gamma$ to experiment, since $\Gamma$ depends on the phase space.  L\"usher's  approach has been verified on several conventional mesonic resonances. 
\item The {\it bound state (B)} in $M_1 M_2$ scattering is realized when $T(p)$ has a pole at $p_B^2<0$ or  $p_B= i|p_B|$
\begin{equation}
\label{B}
T=  \frac{1}{\cot(\delta_l(p_B))-i}=\infty\ , \ \  \cot(\delta(p_B))=i\ ,\quad m_B=E_{H_1}(p_B)+E_{H_2}(p_B)~.  
\end{equation}
 The  location of an s-wave shallow bound state can be obtained by parametrizing $\delta_0$ near threshold and finding $p_B$ which satisfies $\cot(\delta(p_B))=i$.   
 \end{itemize}
 
  \section{Vector and scalar resonances } 
  
  Until recently, all charmonia above open-charm threshold were treated ignoring the strong decay 
  to a pair of charmed mesons. The first exploratory  simulation aimed at determining the masses as well as the decay widths  of these resonances was presented in \cite{Lang:2015sba}.  The Breit-Wigner-type fit of the $D\bar D$ phase shift in $p$-wave leads to the resonance mass  and width of   $\psi(3770)$ (see Fig.  \ref{fig:vector_charmonia} and Table \ref{tab:vector_charmonia}).    The $\psi(2S)$ in Fig.    \ref{fig:vector_charmonia} \label{fig:vector_charmonia} appears as a bound state pole below threshold.

\begin{table}[htb]
\begin{center}
\begin{tabular}{c|cc} 
$\psi(3770)$  & $m_{res}$ [MeV] & $g$ (no unit)\cr
\hline
Lat ($m_\pi=266~$MeV)   &  $3774 \pm 6 \pm 10$ & $19.7\pm 1.4$\cr 
Lat ($m_\pi=156~$MeV)   &  $3789 \pm 68 \pm 10$ & $28\pm 21$\cr 
Exp. & $3773.15\pm 0.33$ & $18.7\pm 1.4$ \cr 
\end{tabular}
\end{center}
\caption{\label{tab:vector_charmonia}  Parameters of the resonance $\psi(3770)$ in $D\bar D$ scattering.  The width is parametrized in terms of the coupling $g$ as   $\Gamma=g^2p^3/(6\pi s)$ \cite{Lang:2015sba}. }
\end{table}

In the scalar channel, only the ground state $\chi_{c0}(1P)$ is understood and there is no commonly accepted candidate  for its first excitation  $\chi_{c0}(2P)$ although PDG assigns  it to  $X(3915)$. The analytic studies \cite{Guo:2012tv,Olsen:2014maa} argue that the $X(3915)$ can probably not be identified with the $\chi_{c0}(2P)$, while  \cite{Zhou:2015uva} finds that the experimental data on $X(3915)$ could actually be  compatible with $J^{PC}=2^{++}$. These studies suggest that a broad structure observed in the $D\bar D$  invariant mass  represents $\chi_{c0}(2P)$.  
The phase shift for $D\bar D$ scattering in $s$-wave was extracted from lattice \cite{Lang:2015sba}, 
but the resulting phase shift  also does not allow a clear answer to the puzzles in this channel.  The lattice data provide an indication for a yet-unobserved narrow resonance slightly below $4~$GeV with $\Gamma[\chi_{c0}^\prime \to D\bar D]$ below $100~$MeV.  A scenario with this narrow resonance and a pole in the $D\bar D$ scattering matrix at $\chi_{c0}(1P)$ agrees  with the energy-dependence of the phase shift \cite{Lang:2015sba}. Three other scenarios  are not supported by the lattice data: just one narrow resonance, just one broad resonance  (proposed in  Guo \& Meissner \cite{Guo:2012tv} and Olsen \cite{Olsen:2014maa}), or one narrow and one broad resonance.    Further experimental and lattice QCD efforts are required to map out the $s$-wave $D\bar D$ scattering in more detail.

\begin{figure}[htb] 
\begin{center}
\includegraphics[width=0.4\textwidth,clip]{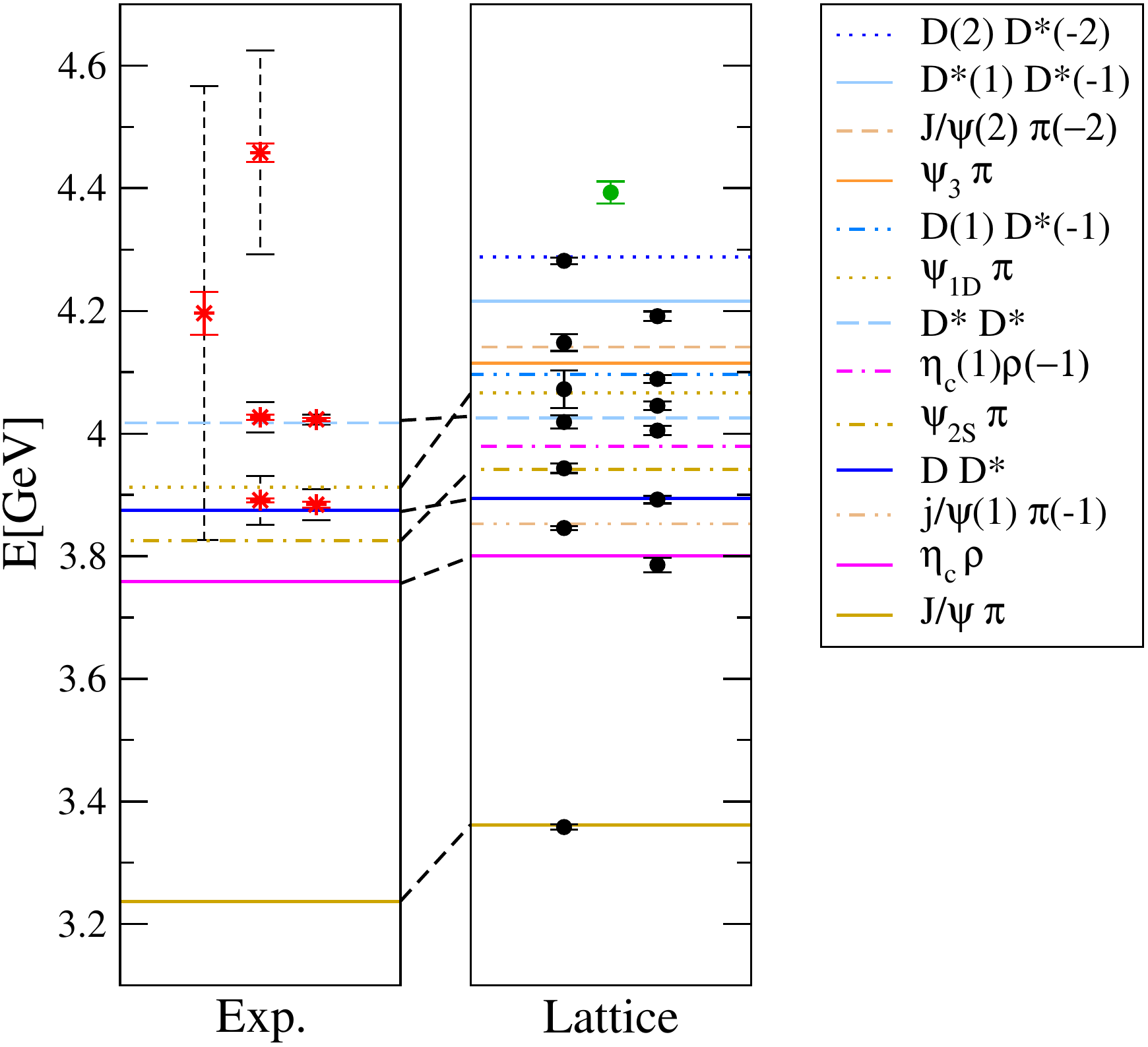}  $\qquad$
\includegraphics[width=0.4\textwidth,clip]{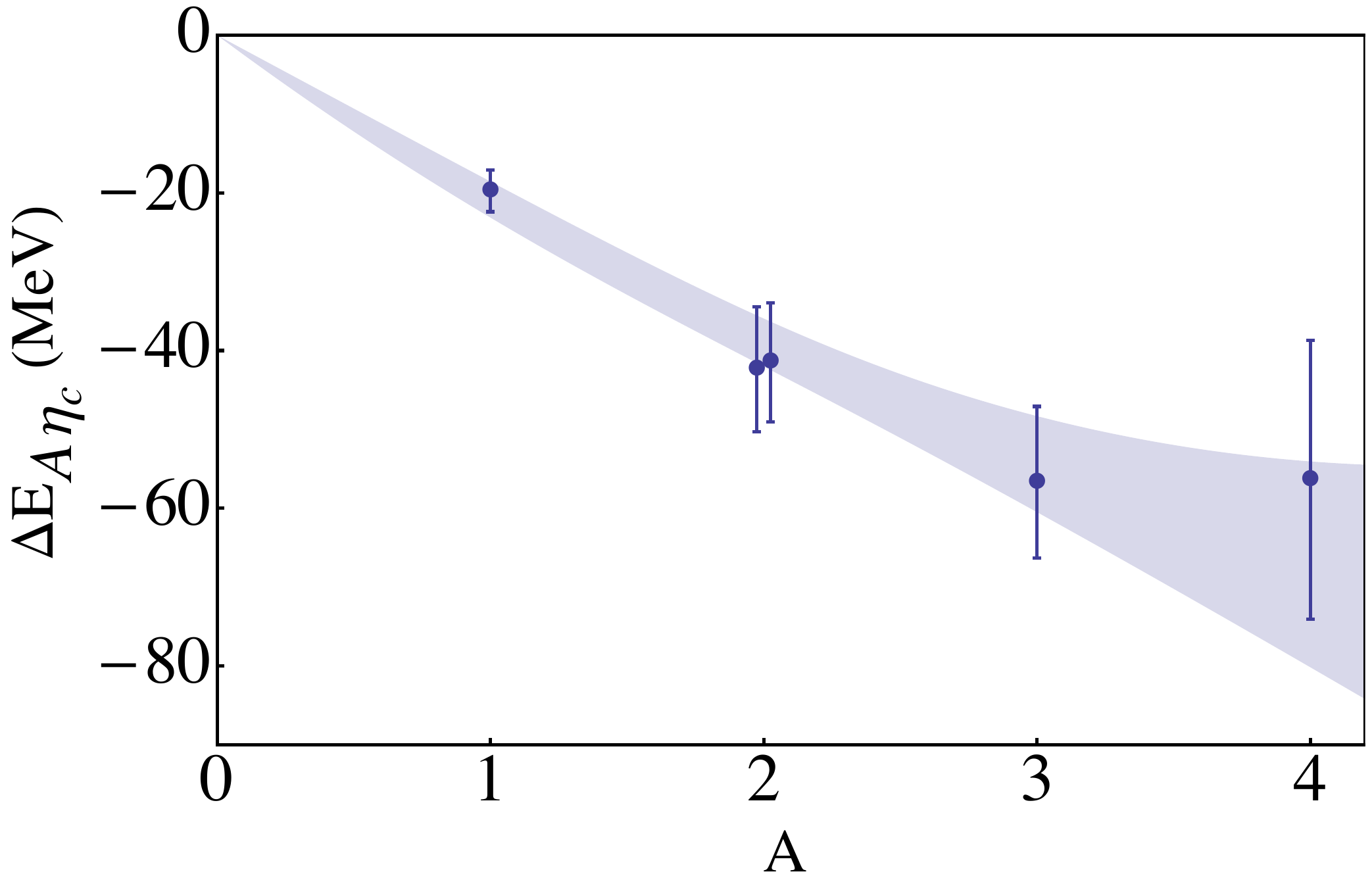}   
\caption{\label{fig:Zc_prelovsek}   Left: Energies of eigenstates in $Z_c^+$ channel from  simulation \cite{Prelovsek:2014swa} are given by black circles. They correspond to the expected two-meson states, whose non-interacting energies are given by the horizontal lines. Right: Binding energies of $\eta_c$ and nuclei with $A$ nucleons from \cite{Beane:2014sda}. The bound state at $A=1$ is a pentaquark candidate $\eta_c N$. }
\end{center}
\end{figure}

\section{Charmonium-like states $X(3872)$ and $Y(4140)$}

The $X(3872)$ lies  experimentally on $D^0\bar D^{0*}$ threshold and its existence on the lattice can not be established without taking into account the effect of this threshold. This was first done by simulating $D\bar D^*$ scattering  in \cite{Prelovsek:2013cra}, where   a pole in $D\bar D^*$ scattering matrix was found just below the threshold in $I(J^{PC})=0(1^{++})$ channel  (\ref{B}). The pole is associated with a bound state $X(3872)$ and its location is shown in Fig. \ref{fig:vector_charmonia}. The more recent simulation using HISQ action confirms the existence of the pole just below the threshold \cite{Lee:2014uta,Lee:2014uta_private}.

The lattice study \cite{Padmanath:2015era} investigated which Fock components are essential for appearance of $X(3872)$ with $I=0$ on the lattice. The energy eigenstate  related to $X(3872)$ appears in the simulation only if  $D\bar D^*$ as well as $\bar cc$ interpolating fields are employed.  The $X(3872)$ does not appear in absence of  $\bar cc$ interpolators, even if  (localized)  interpolators $[\bar c\bar q]_{3_c}[cq]_{\bar 3_c}$ or $[\bar c\bar q]_{6_c}[cq]_{\bar 6_c}$  are in the interpolator basis. This indicates that $\bar cc$ Fock component is most likely more essential for $X(3872)$ than the diquark-antidiquark one. 

The search for charged $X(3872)$ was performed in $J^{PC}=1^{++}$ channel with $D\bar D^*$, $J/\psi \rho$, $[\bar c\bar d]_{3_c}[cu]_{\bar 3_c}$ and  $[\bar c\bar d]_{6_c}[cu]_{\bar 6_c}$ interpolators and no candidate was found \cite{Padmanath:2015era}. The reliable search for the neutral $I=1$ state would need to incorporate isospin breaking effects \cite{Garzon:2013uwa}, but that has not been performed on the lattice yet.  

The experimental candidate $Y(4140)$ with hidden strangeness was observed in $J/\psi \phi$ invariant mass and has unknown $J^P$ at present. 
The lattice search was performed in $J^{PC}=1^{++}$ channel with $D_s\bar D_s^*$, $J/\psi \phi$, $[\bar c\bar s][cs]$  interpolators and no candidate was found \cite{Padmanath:2015era}. The s-wave and p-wave $J/\psi\;\phi$ scattering phase shift from \cite{Ozaki:2012ce} do not support the resonant structure.  

 \begin{figure}[htb] 
\begin{center}
\includegraphics[width=0.32\textwidth,clip]{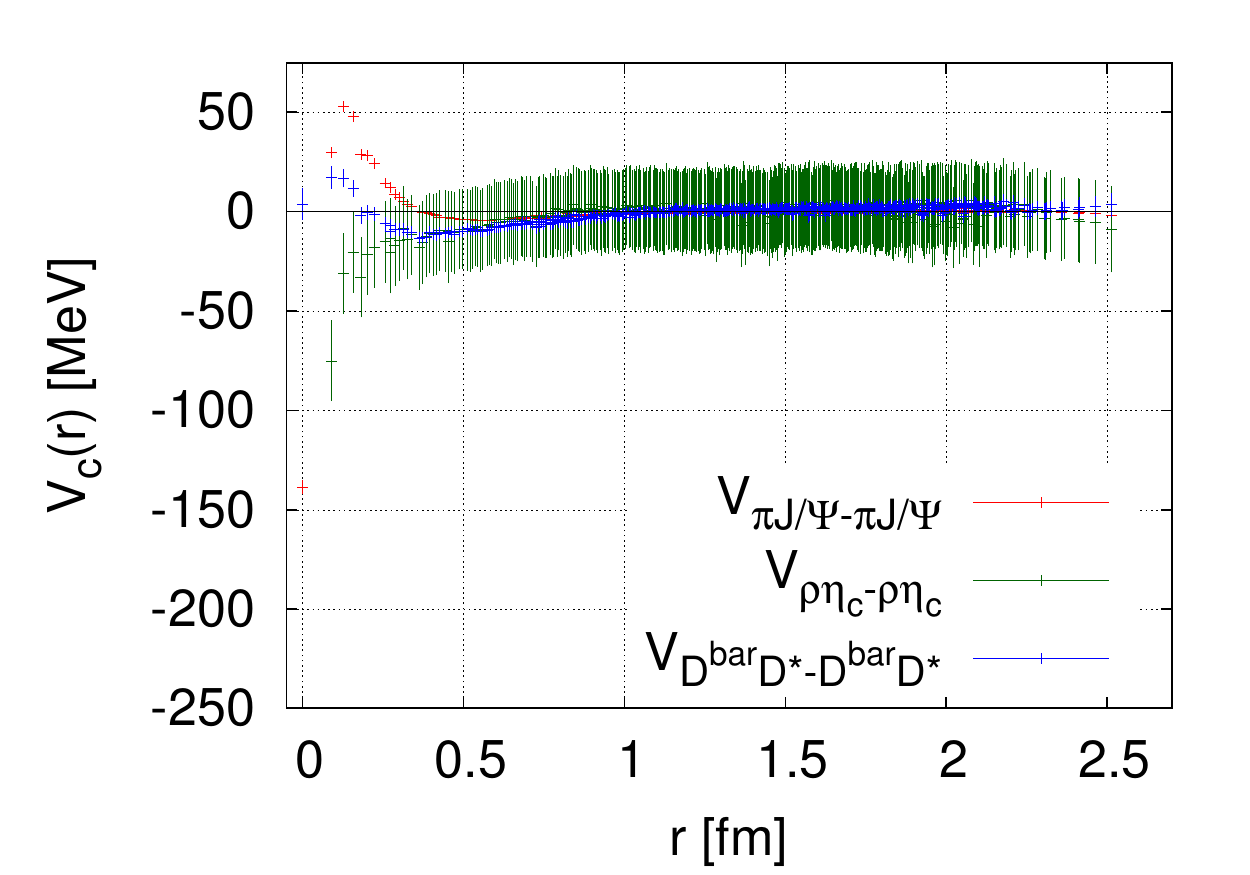}  
\includegraphics[width=0.32\textwidth,clip]{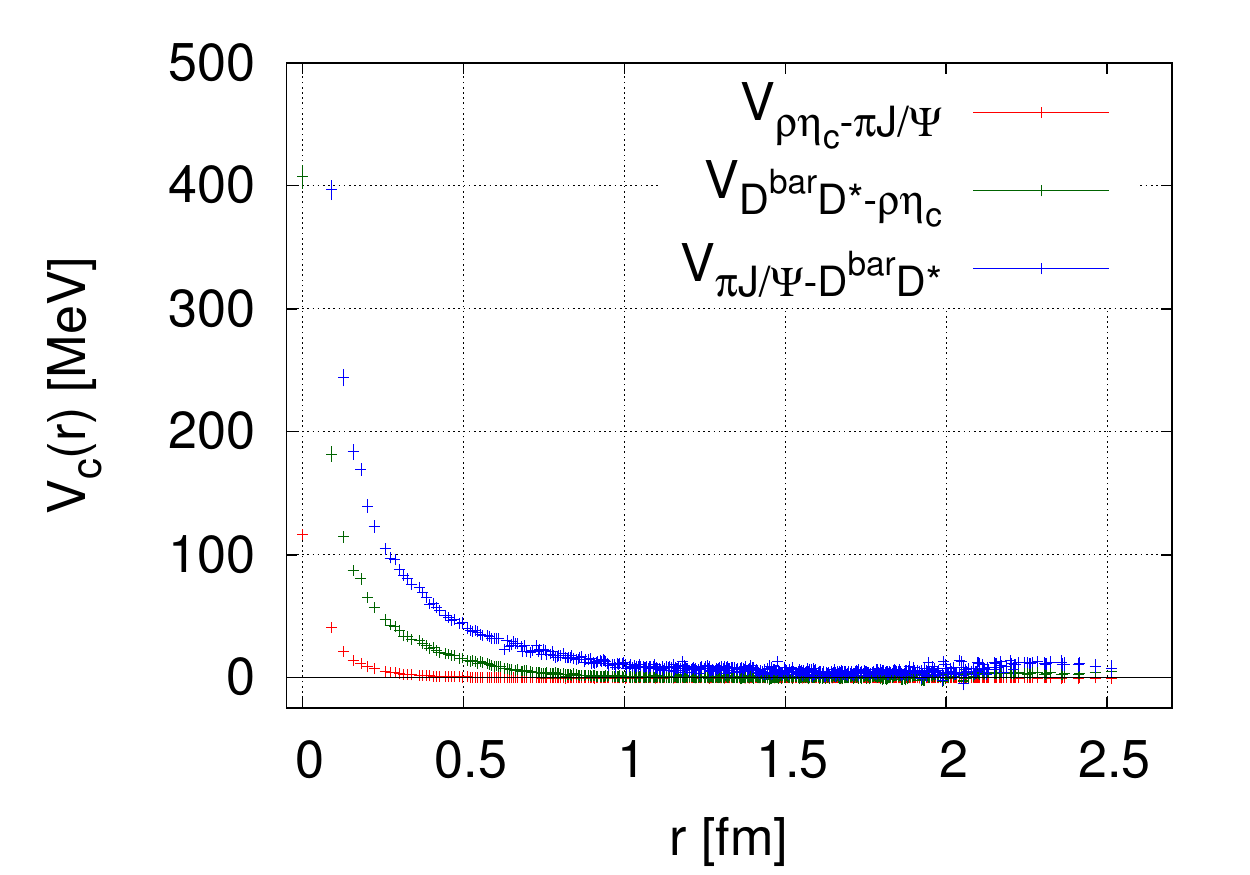}  
\includegraphics[width=0.32\textwidth,clip]{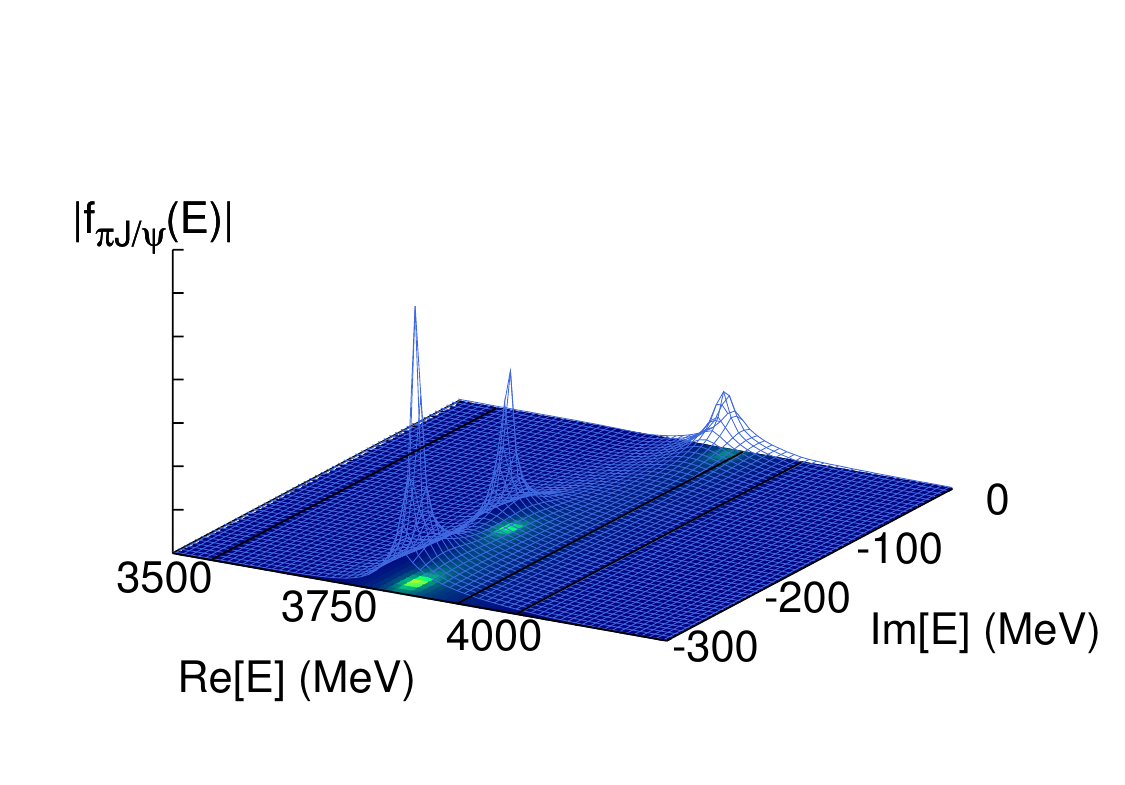}  
\includegraphics[width=0.38\textwidth,clip]{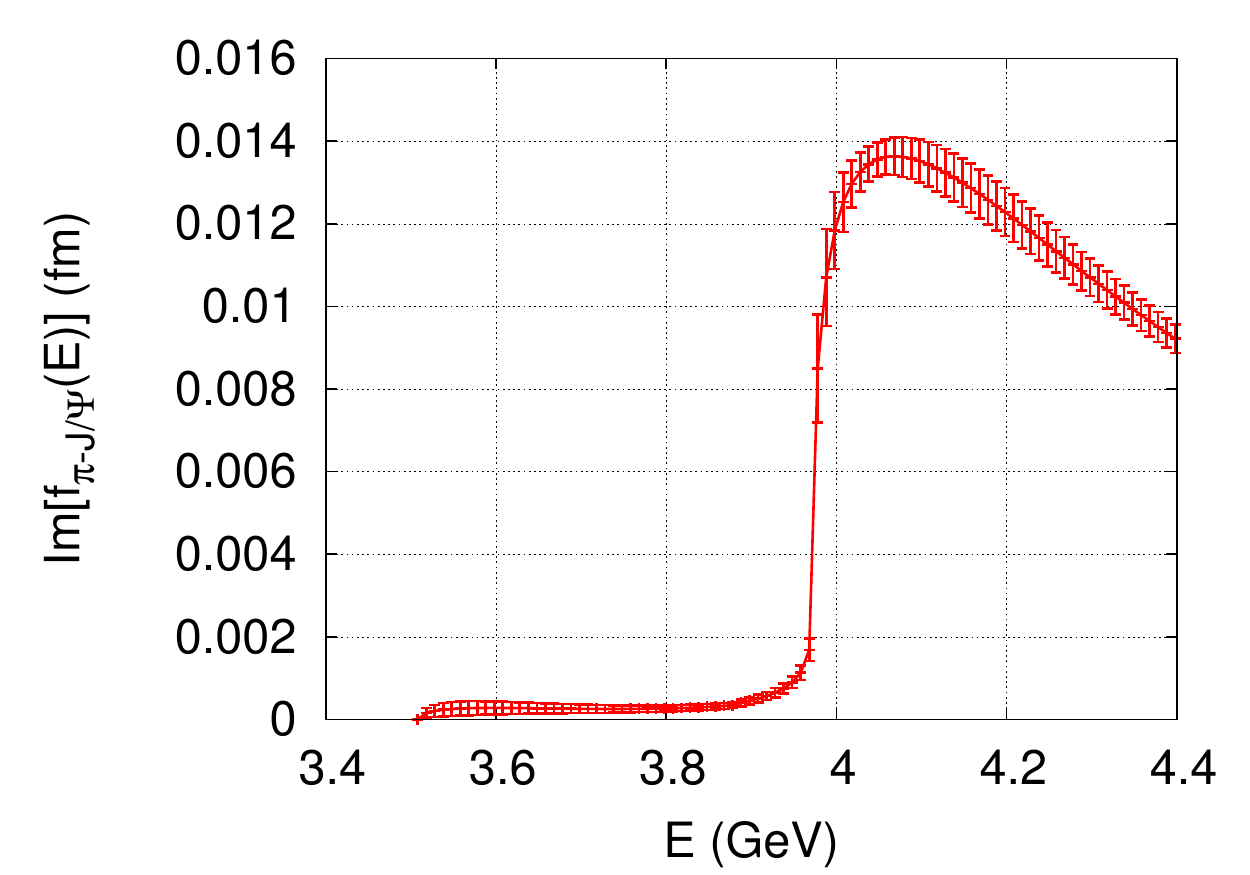}  $\quad$
\includegraphics[width=0.36\textwidth,clip]{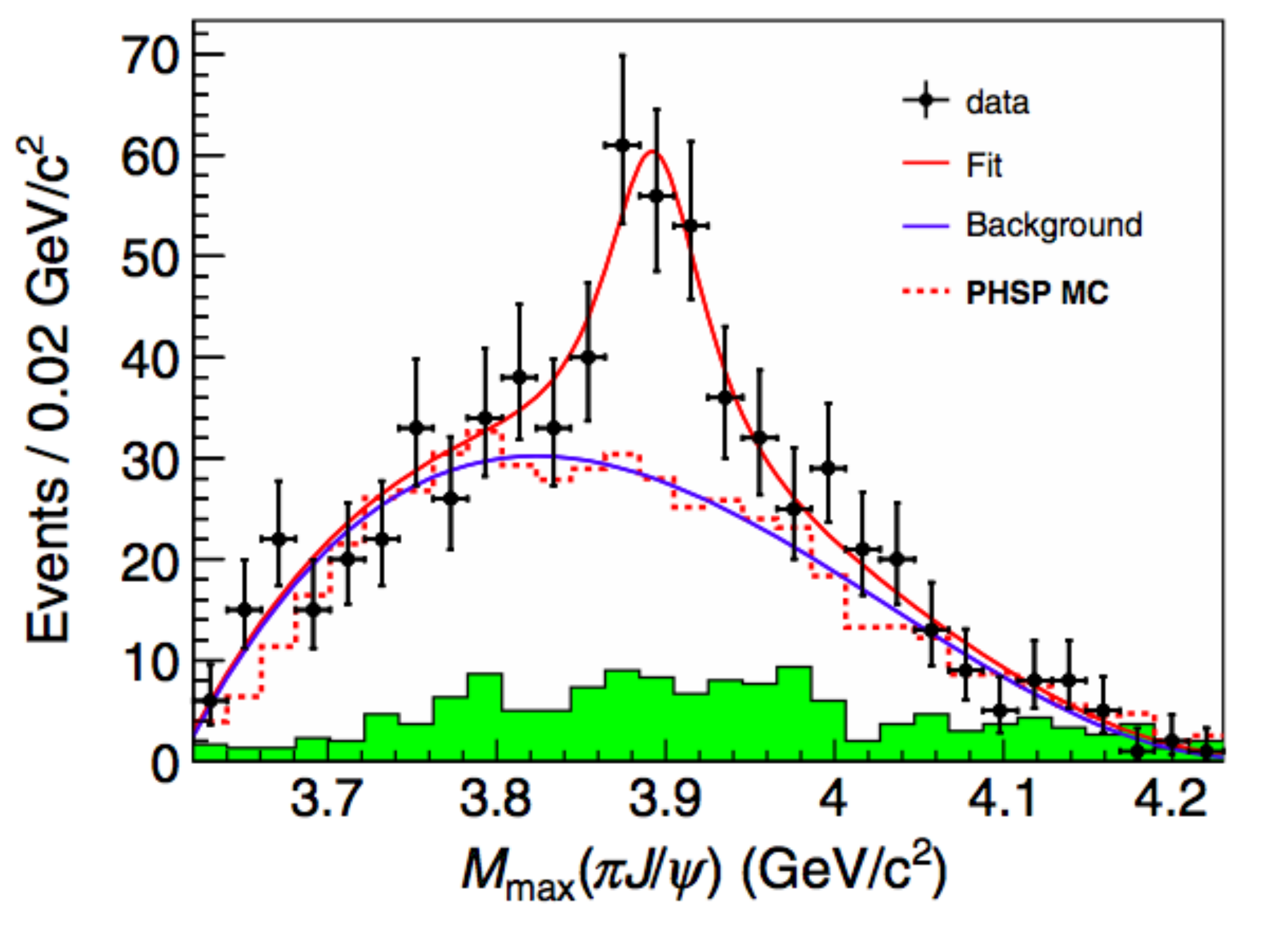}  
\includegraphics[width=0.38\textwidth,clip]{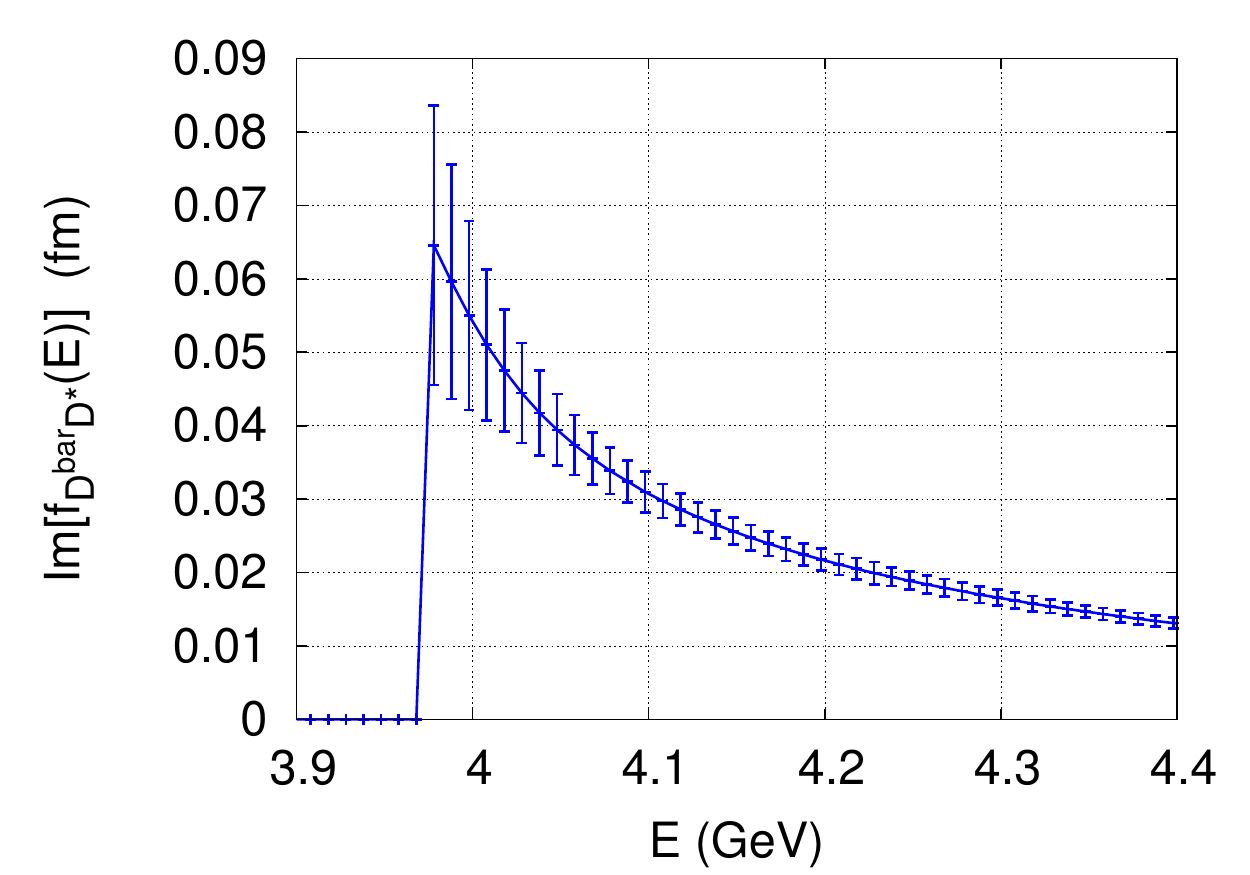}  $\quad$
\includegraphics[width=0.36\textwidth,clip]{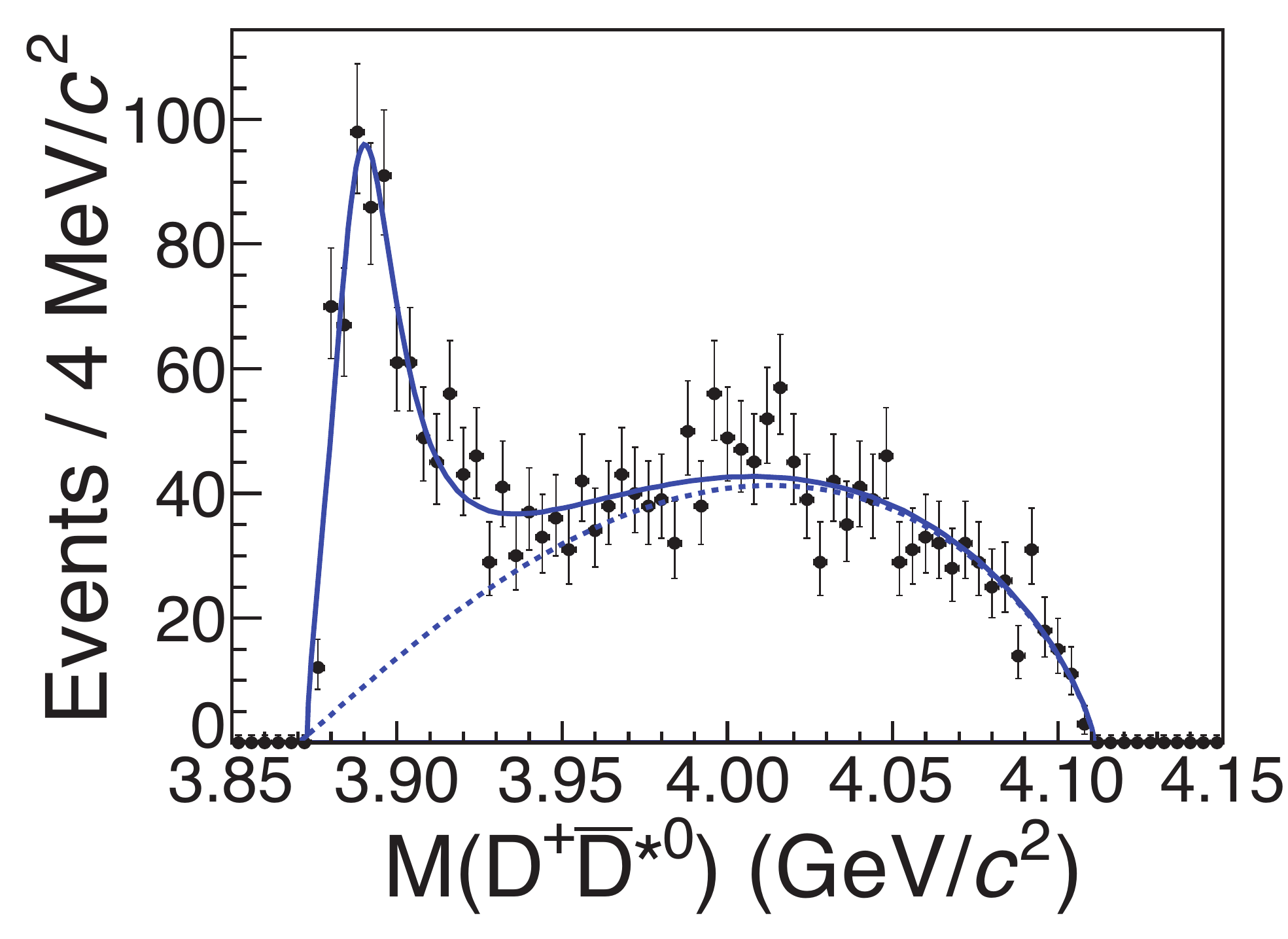}  
\caption{\label{fig:Zc_halqcd}   The HALQCD results related to $Z_c^+(3900)$ where scattering amplitude $f$ is defined through $d\sigma/d\Omega=|f(E)|^2$ \cite{halqcd:tmp}. Top left: potentials $V(r)$ for three channels. Top middle: crossed-channel potentials $V(r)$ with sizeable $J/\psi\pi - D\bar D^*$.   Top right: Poles of the scattering matrix in the complex energy plane.  The quantity related to the number of particles  $Im[f_i]\propto j_i \sigma_i$ for channels $i=J/\psi \pi,~D\bar D^*$ compared to the experimental shapes for $J/\psi \pi$ \cite{Liu:2013dau} and $D\bar D^*$  \cite{Ablikim:2013xfr}.}
\end{center}
\end{figure}

\section{ Charged charmonium-like states $Z_c$}

The lattice search for the manifestly exotic states $Z_c^+$ with flavour content $\bar cc\bar du$ and $I^G(J^{PC})=1^+(1^{+-})$ is very challenging since 
the experimental candidates lie above several thresholds and can decay in several final states via strong interaction. The reliable treatment requires simulation of coupled channels and extraction of coupled channel scattering matrix. This has been done using the L\"uschers method  only for one system  in the light sector \cite{Dudek:2014qha}. 

  The lattice search for $Z_c^+$ in the energy region below $4.3~$GeV \cite{Prelovsek:2014swa} has therefore been performed using a simplified approach. The challenge is that there are 13 two-meson eigenstates ($J/\psi \pi$, $D\bar D^*$, $\eta_c \rho$, ...) with $I^G(J^{PC})=1^+(1^{+-})$ in this energy region for the $L\simeq 2~$fm, as shown  by horizontal lines in Fig. \ref{fig:Zc_prelovsek}.  The explicit simulation of all these coupled channels (with a large number of   meson-meson and diquark-antidiquark interpolators)  renders 13 expected two-meson eigenstates in the region (shown by black circles). No additional energy eigenstate has been found, therefore no candidate for $Z_c^+$. This has been recently confirmed also by a simulation using HISQ action \cite{Lee:2014uta}. 
  Note that an extra eigenstate (in addition to expected two-meson states)  has been found for all resonances and all bound states that have been established on the lattice up to now (this implies, for example,  also for the bound state $X(3872)$ and resonance $\psi(3770)$ discussed in this talk).  The absence of an additional eigenstate for $Z_c(3900)$  could indicate that this experimental state is not related to a conventional resonance pole, but could be a manifestation of coupled channel effect. 
  
  This possibility was investigated within the so-called HALQCD approach \cite{HALQCD:2012aa} to extract the scattering matrix of the coupled channels $D\bar D^*$, $J/\psi \pi$ and  $\eta_c \rho$  \cite{halqcd:tmp}. The potential $V_{\pi J/\psi -\pi J/\psi}(r)$ related to Nambu-Bethe-Salpeter equation is determined between the $J/\psi$ and $\pi$  as a function of their separation $r$. This is presented in   Fig. \ref{fig:Zc_halqcd} together 
  with potential for the other two channels, as well as potential corresponding to coupling between different channels according to coupled-channel HALQCD formalism \cite{Aoki:2012bb}.  The off-diagonal potential between channels $\pi J/\psi$ and $D\bar D^*$ is larger than other potentials, which   seems to indicate a sizeable coupled channel effect near $D\bar D^*$ threshold. The $3\times 3$ matrix of potentials renders $3\times 3$ scattering matrix\cite{Aoki:2012bb}. This in turn gives a quantity related to the number of events $N_i \propto \sigma_{i} j_i $ in each channel $i$ as a function of invariant mass, which is compared to experiment  in Figure \ref{fig:Zc_halqcd}. Both $N_{\pi J/\psi}$ and $N_{D\bar D^*}$ show  enhanced peaks above $D\bar D^*$ threshold that resemble experimental line shapes for $Z_c^+(3900)$.   The pole structure that emerges from the extracted scattering matrix is shown in the complex energy plane in Fig. \ref{fig:Zc_halqcd}. Two poles are found at $Im[E_{cm}]<0$ and   $Re[E_{cm}]<m_{D}+m_{\bar D^*}$, while a pole would be expected at  $Re[E_{cm}]>m_{D}+m_{\bar D^*}$ if $Z_c(3900)$ would be conventional resonance. The authors conclude that $Z_c^+(3900)$ most likely does not correspond to the conventional resonant pole, while the cross section peak is related to sizeable $\pi J/\psi - D\bar D^*$ coupled channel effect.   It remains to be explored whether the extracted scattering matrix is in agreement with the absence of an additional energy eigenstate near $3.9~$GeV in  \cite{Prelovsek:2014swa,Lee:2014uta}. The conclusions based on the HALQCD approach need to be verified  using the L\"uscher-type approach, which has been shown to render reliable rigorous results for several conventional meson resonances.  The HALQCD approach has not been thoroughly tested  on the conventional meson resonances yet. 
  
  The s-wave and p-wave phase shifts  near $D\bar D^*$ threshold were determined using only $D\bar D^*$ interpolating fields in    \cite{Chen:2014afa}, which may not be reliable since the ground state of the system is $J/\psi \pi$.  The authors conclude that  no evidence for $Z_c^+(3900)$ is found.

\section{Pentaquarks and $\bar cc$-nucleus bound states}

The NPLQCD collaboration finds an interesting evidence for a $\eta_c N$ bound state approximately $20~$MeV below $\eta_c N$ threshold,  as shown in Fig.  \ref{fig:Zc_prelovsek}  \cite{Beane:2014sda} ($N$ denotes nucleon). To my knowledge, this is the only pentaquark candidate containing $\bar cc$ from lattice studies up to now. As the simulation is done at $SU(3)$ flavour symmetric point corresponding to $m_\pi\simeq 800~$MeV, it is not clear yet whether this bound state persists to physical $m_\pi$.  LHCb has recently found two pentaquark  resonances   in $J/\psi p$, located about $400~$MeV above threshold \cite{Aaij:2015tga}. The lattice simulation of those is much more challenging due to several open channels and has not been performed yet. 

 Bound states of $\eta_c$ and nuclei ware also found in  \cite{Beane:2014sda}, where binding energies are given in Fig. \ref{fig:Zc_prelovsek}.  
 
 \begin{figure}[htb]
\centering
\includegraphics[width=0.44\textwidth,clip]{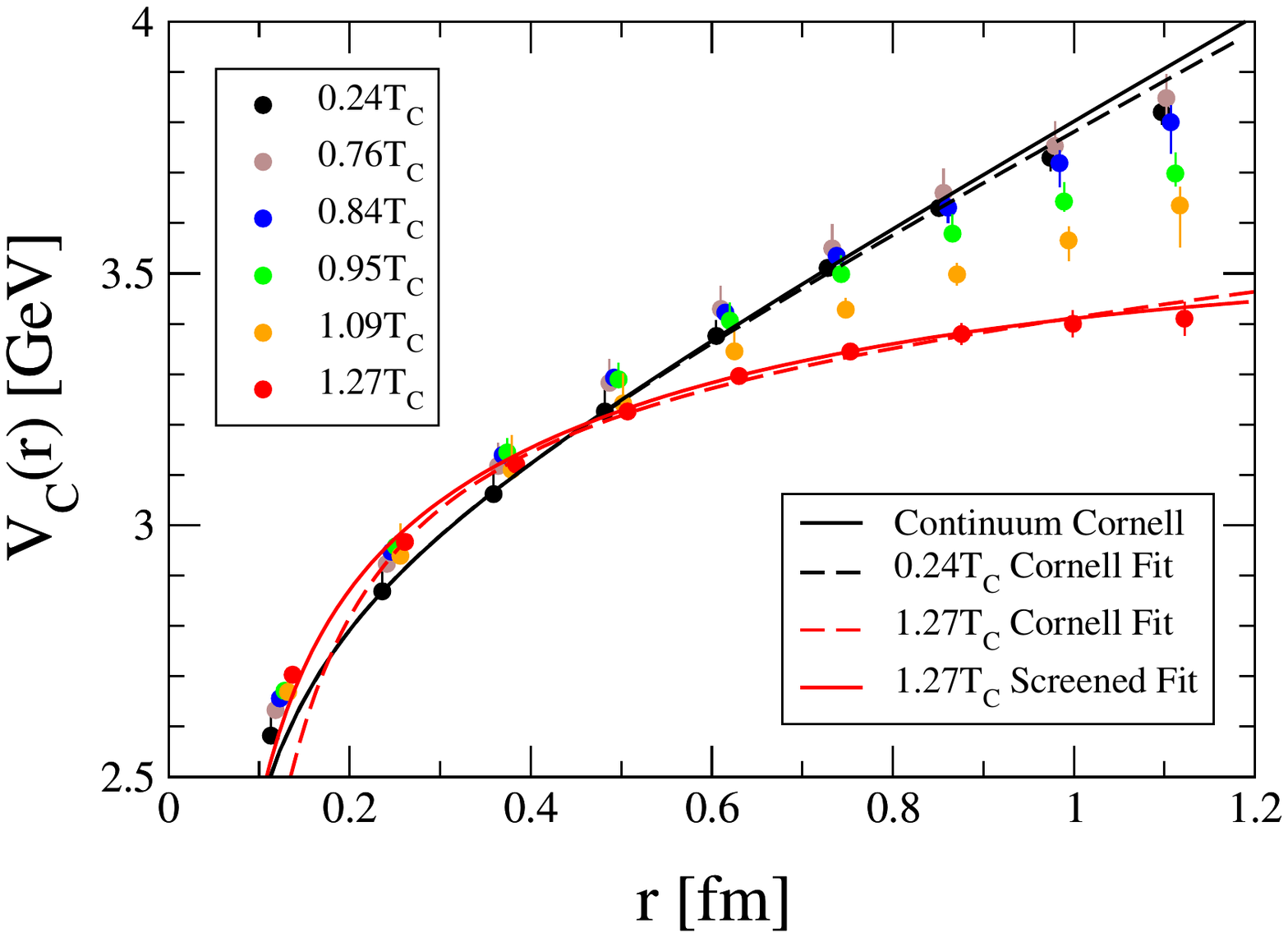} 
\includegraphics[width=0.44\textwidth,clip]{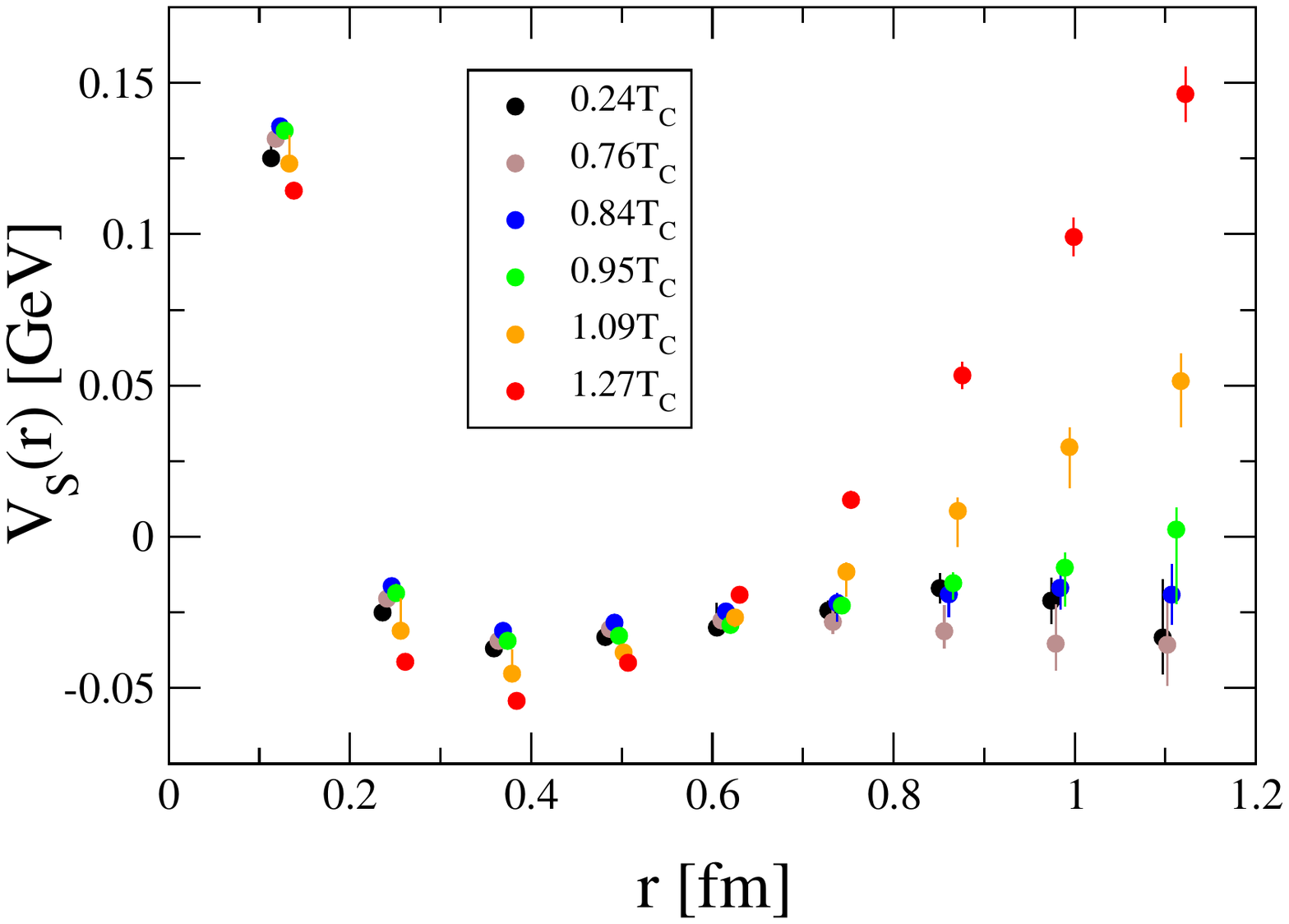} 

\vspace{-4.5cm}

 \caption{ The central and spin dependent potential $V(r)=V_C(r)+V_S(r) \vec s_1 \cdot \vec s_2$ between $\bar c$ and $c$ for various temperatures  \cite{Allton:2015ora}.  }
\label{fig:V}
\end{figure}
 
\section{Charmonium potential at nonzero temperature}

The potential $V(r)$ between $c$ and $\bar c$ was extracted using HALQCD method \cite{HALQCD:2012aa} at a range of finite temperatures $T$   \cite{Allton:2015ora}. The central and spin dependent parts $V(r)=V_C(r)+V_S(r) \vec s_1 \cdot \vec s_2$ are presented in Fig. \ref{fig:V}. The lines in $V_C$ present  Cornell and Debye screening fit of the results, and the resulting Debye screening mass acquires a nonzero value roughly at $T_c$.  The spin-dependent part has a repulsive core which roughly resembles the familiar parametrisation with the $\delta(r)$ function. 
 
\section{Summary}

 Recent lattice QCD studies of charmonium and charmonium-like states were reviewed.   The main challenge for the future lattice simulations is the extraction of the scattering matrix relevant to the experimentally interesting states. This will, for example,  involve two or more coupled channels for tetraquarks $Z_c^+$ or pentaquarks $P_c$.   
  
\Acknowledgements I would like to thank  C. Allton, C. DeTar, Y. Ikeda,  S.-H. Lee, M. Savage and Yi-Bo Yang  for sending me material in preparation for this talk, and Y. Ikeda, C. B. Lang and D. Mohler for reading the manuscript.  I acknowledge the support from Slovenian Resarch Agency ARRS project N1-0020,  Austrian Science Fund FWF I1313-N27 and U.S. Department of Energy contract DE-AC05-06OR23177 under which Jefferson Science Associates operates Jefferson Laboratory.


\bibliographystyle{h-physrev4}
\bibliography{Lgt_charm15}

\end{document}